\documentclass[10pt]{iopart}
\usepackage{graphicx}
\usepackage{float}
\begin{document}

\title[ ]{The ElectroHydroDynamic force distribution in surface AC Dielectric Barrier Discharge actuators: do streamers dictate the ionic wind profiles?}
\author{K Kourtzanidis\footnote{Present address: CERTH - Centre for Research and Technology Hellas, 57001 Thermi, Thessaloniki, Greece}, G Dufour, F Rogier}%
\address{ONERA - The French Aerospace Lab, 31000, Toulouse, France}%



\ead{kourtzanidis@certh.gr}

\begin{abstract}
We show that the spatio-temporal ElectroHydroDynamic (EHD) force production in surface AC-Dielectric Barrier Discharge (AC-DBD) actuators is strongly influenced by both the streamer regime during the positive phase and the micro-discharge regime during the negative phase. Focusing on the spatial EHD force profiles, we demonstrate that the ionic wind spatial distribution can only be explained by the positive contribution of the streamer regime. The location of maximum ionic wind is found to be directly linked with the maximum elongation of the streamers at several millimeters from the exposed electrode. In both positive and negative phases of the AC-DBD operation, residual volumetric and surface charges once again linked to the streamer formation and afterburn, result to a variety of positive EHD force zones which, when time-averaged in one AC period, contribute to the generation of the experimentally observed induced thin wall jet. Through a thorough elaboration of our numerical results, we provide an illustrative explanation of the EHD force spatio-temporal evolution, showcase the importance of streamers and retrieve a correct representation of the ionic wind spatial profiles when compared to experiments.

\end{abstract}
\maketitle

\section{Introduction and methods}
ElectroHydroDynamic (EHD) flows induced by surface Alternative Current Dielectric Barrier Discharge (AC-DBD) actuators have found use in a variety of applications mainly as means of aerodynamic flow control~\cite{moreau2007airflow}. Separation delay and flow re-attachment, turbulent enhancement and laminar-to-turbulent transition control, vortex generation, turbine blades aerodynamic enhancement are a few of such applications. Despite the numerous studies, the spatio-temporal distribution of the induced EHD flow or ionic wind is yet to be fully understood.  Experiments have demonstrated that the plasma discharge nature is very different in both half-phases~\cite{benard2014electrical}. In the positive going cycle, high current streamer discharges form above the dielectric layer while in the negative going cycle, lower current but higher frequency micro-discharges are present. The contribution of each phase to the EHD force production and consequent ionic wind profiles is a controversial subject. Concerning temporal aspects, push-push and push-pull scenarios have been proposed and supported by experiments and simulations~\cite{kuhnhenn2016interrelation, benard2013time}. Concerning spatial aspects, on one hand experimental studies have been used to retrieve the force distribution~\cite{benard2013time, kotsonis2011measurement, enloe2008time, kriegseis2012piv, kriegseis2013velocity}. These studies are mostly based on strong assumptions (pressure gradients, local acceleration, turbulent fluctuations) and an inverse Navier-Stokes (NS) procedure that render the results ambiguous as they do not correlate with velocity measurements. On the other hand, numerical and theoretical studies on the ionic wind profiles show in general good agreement with experiments in terms of overall thrust production but fail to capture the maxima position of the induced wall-jet. Most of these numerical studies suffer from low accuracy due to the numerical schemes used and simplifying assumptions made and/or insufficient spatial discretization while limited information on the spatial distribution of the EHD force has been provided (e.g. Ref.~\cite{boeuf2007electrohydrodynamic}, see also references in Ref.~\cite{jointpaper}).

In this work, based on the detailed numerical simulations of our recently published work in Ref.~\cite{jointpaper}, we answer two important questions on the spatio-temporal incertitude of EHD force and ionic wind produced by surface AC-DBD discharges :
Why the ionic wind spatial profiles present maxima at a distance of several millimeters from the exposed electrode as observed experimentally~\cite{zhang2020flow, debien2012unsteady, durscher2012evaluation, kotsonis2012performance}? How does each phase of the AC cycle contribute to the induced flow (magnitude and direction)? To do so, we build and elaborate on the results of our recently published work in Ref.~\cite{jointpaper}, demonstrate the time-averaged spatial EHD force and ionic wind produced by an AC-DBD actuator and propose an illustrative explanation of the complex plasma-flow interaction. A self-consistent modeling approach has been followed in order to obtain the full cycle characteristics of the surface AC-DBD operation. Details on the numerical and physical models used as well as discussion on the main assumptions made can be found in Ref.~\cite{dufour2015numerical, jointpaper}. The AC-DBD under study is a 100 kHz, 20 kV actuator.

\section{Time-dependent plasma characteristics and EHD force production zones during an AC cycle }
 In Ref.~\cite{jointpaper}, we show that the AC-DBD operation can be decomposed in two phases which are nevertheless strongly inter-connected in agreement with several experimental studies~\cite{moreau2007airflow, benard2014electrical}. In summary our findings demonstrate that: In the positive phase, a positive corona-like discharge forms at the active electrode interrupted by a high current surface streamer discharge. The streamer propagates detached from the dielectric surface acting as a virtual anode. When its propagation is stopped, it slowly relaxes and positive ions from its body charge the dielectric surface contributing to an elongated zone of positive potential and consequent high electric field at a distance of several millimeters from the active electrode. In the negative phase, volumetric charge separation leads to the initiation of repetitive microdischarges which attach to the active electrode (cathode in this phase) forming a thin cathode layer. Each microdischarge terminates with the propagation of a thin plasma layer attached to the dielectric surface. The positively charged portion of the dielectric which persists from the positive/streamer phase, pulls this layer further and further until it is quenched due to electrons and negative ions which drift towards the dielectric. In the relaxation phase between consecutive microdischarges, positive ions are repelled outwards from the dielectric surface and the thin ion layer attached to the dielectrict consists of mainly negative ions. \\

\begin{figure}[h]
\centering
\includegraphics[width=0.7\linewidth]{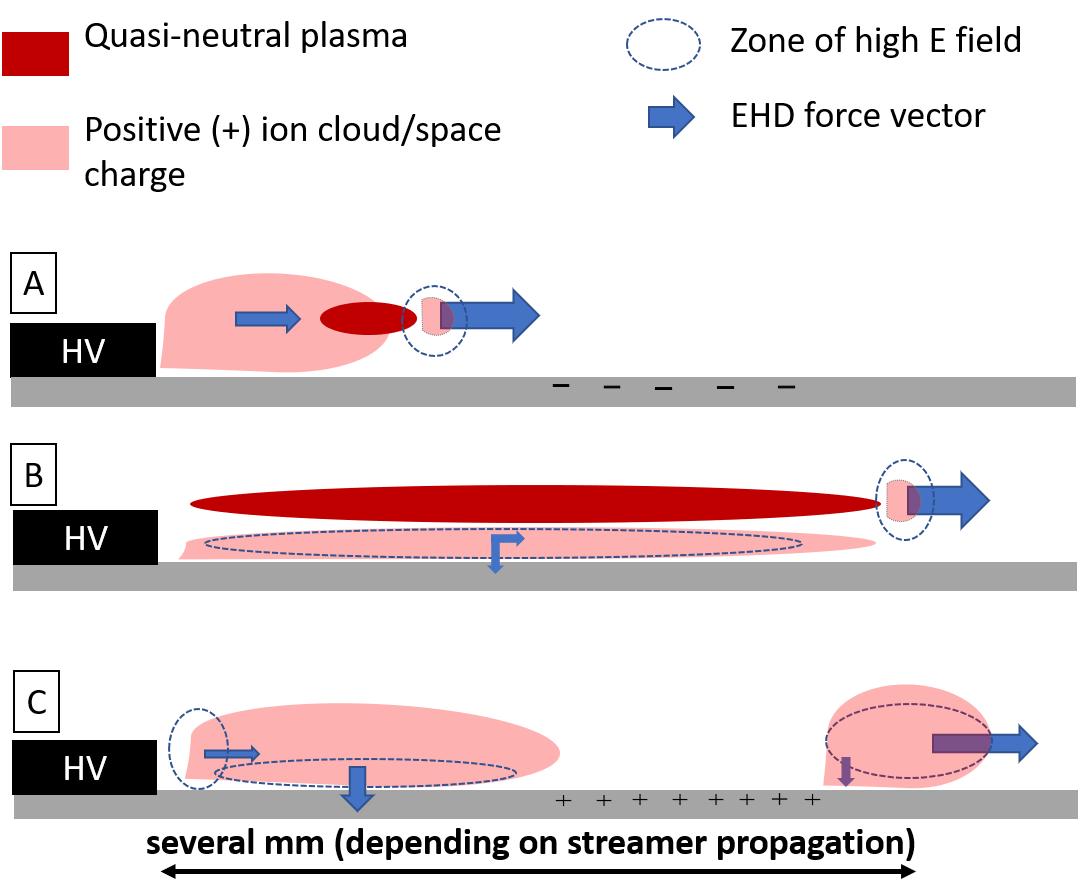}
\caption{Illustration of the positive phase EHD force production zones: A) Initially the positive ion cloud expands over the dielectric surface until space charge effects initiate the quasi-neutral streamer. A positive space charge and high electric field region exists at the head of the streamer. B) The streamer propagates quickly parallel to the dielectric surface. In addition to the space charge at its head, a zone of high electric field populated by diffused positive ions exist as a sheath between the streamer body and the dielectric. C) At the relaxation phase, the streamer relaxes and charges the dielectric positively. The virtual anode formation due to the streamer (after-burn) and dielectric charging (relaxation) enhances the electric field at a distance of several millimeters and leads to an elongated zone of EHD force production. The analysis is based on our detailed simulations of Ref.~\cite{jointpaper}.}
 \label{fig:demo_positive}
 \end{figure}

\begin{figure}[h]
\centering
\includegraphics[width=0.7\linewidth]{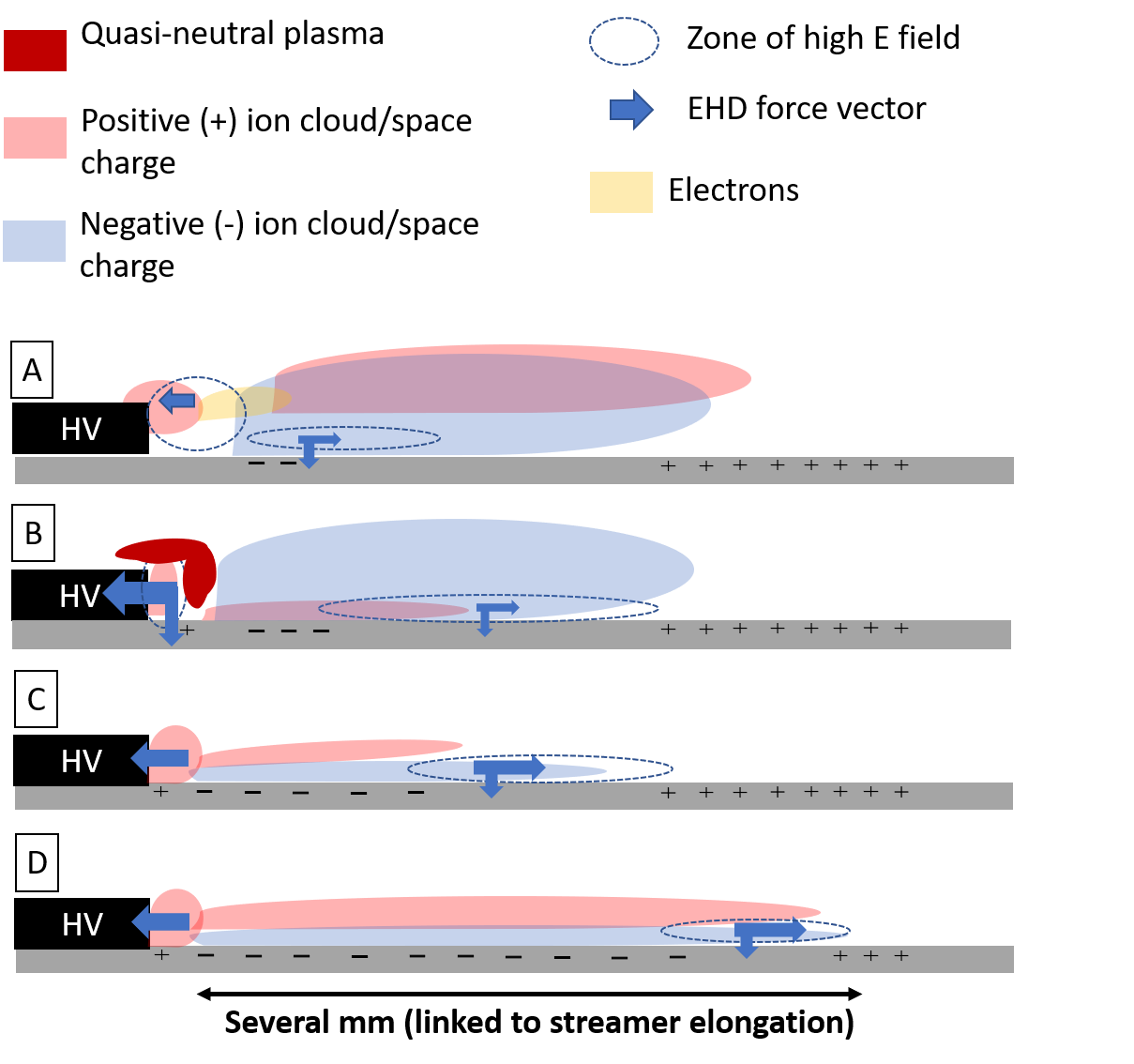}
\caption{Illustration of the negative phase EHD force production zones: A) Electrons produced near the active electrode drift towards the dielectric charging it negatively. Positive ions near the cathode contribute to a negative EHD force. The negative ion cloud region near the dielectric (remnant from the positive phase) produces positive EHD force under the influence of the electric field due to the potential difference between the negative and positive charged portions of the dielectric surface.  B) A quasi-neutral micro-discharge forms and rapidly attaches to the exposed electrode forming a cathode layer. The cathode layer holds very high electric fields, positively dominated space charge and a negative EHD forcing zone appears. Positive ions are generated in the near surface region due to the previously mentioned electric field. C) Once the micro-discharge relaxes and the plasma layer propagates on the dielectric surface, positive ions are repelled from the dielectric leaving a negative ion layer behind. The EHD force is there positive and dominant due to the time-scale of the relaxation phase between each microdischarge. D) The surface ion layer expands after each microdischarge until it reaches the end of the positively charged portion of the dielectric (linked to the streamer elongation during the positive phase). The analysis is based on our detailed simulations of Ref.~\cite{jointpaper}.}
 \label{fig:demo_negative}
 \end{figure}

Based on the aformentioned discharge characteristics, we can identify the EHD force production zones during the two-phase operation of the AC-DBD actuator. In Fig.~\ref{fig:demo_positive} and Fig.~\ref{fig:demo_negative} we present schematically the EHD force production zones in the positive and negative phase respectively. We note that dimensions are not in scale and the representation is illustrative: not all ion cloud zones are presented but only the most important for the EHD force production. The instantaneous EHD force vectors presented in Fig. 4, 9 and 15 of Ref.~\cite{jointpaper} along with the detailed operational description therein help us construct the discussion presented below.  The instantaneous EHD force per unit volume is given by~\cite{boeuf2005electrohydrodynamic}: 
\begin{equation} 
\vec{F}{_{EHD}}=q(n_{+}-n_{-}-n_{e})\vec{E}
\end{equation}
where $q$ is the elementary charge, $n_{+}, n_{-}$ and $n_{e}$ are the positive, negative ions and electrons density respectively and $\vec{E}$ is the electric field vector. It is clear that the EHD force is generated in regions with high electric fields, without charge neutrality and high unipolar charge concentration. 
 
 During the positive phase, the EHD force is located inside 3 main regions: First, the positive ion cloud expanding over the dielectric. Second, the streamer head and the streamer sheath region between its body and the dielectric surface during its short-term propagation. Third, an important part of the EHD force is located in the zone ahead of the streamer maximum elongation length during the relaxation phase. The latter is due to the conductive nature of the streamer and the positive dielectric charging during the (long-term) relaxation phase of the streamer discharge. Both of these factors lead to a zone of enhanced electric field just downstream the streamer body, promoting ionization, positive-ion production which along with the diffusion of the latter from the streamer body contribute to a positive x-directed EHD forcing, as the positive voltage phase persists. The x-directed component is positive in all three regions while a negative region of y-directed force exists in the sheath region between the streamer body and the dielectric. Phase C as illustrated in Fig.~\ref{fig:demo_positive} contributes the most to the EHD force as it lasts several 100s of ns.
 
 During the negative phase, the EHD force is located inside two main regions: First, the negative ion cloud as a remnant of the positive phase streamer with an important x-directed positive component inside a region near the dielectric and between the charged dielectric portions. Second, inside the cathode sheath layer formed due to each micro-discharge generation. The force there is strongly negative and mainly x-directed as positive ions dominate. Third, during the relaxation phase between each microdischarge inside the thin negative ion layer attached to the dielectric. This layer expands further and further after each microdischarge pulse as the electric field between the surface charged regions progresses along. By the end of the negative phase, the region once covered by the streamer discharge is now covered by ion clouds and the thin negative layer near the dielectric which is now negatively charged all along. In all phases, dielectric charging plays an important role in both the discharge behavior as well as the electric field enhancement in critical regions for EHD force production. The reader may find additional details in the captions of Fig.~\ref{fig:demo_positive} and Fig.~\ref{fig:demo_negative}.

In the following section, the spatiotemporal distribution of the EHD force is presented and analyzed, in order to validate the analysis presented above along with the ionic wind spatial profiles obtained through CFD simulations. 

\section{Results}
\subsection{EHD force distribution - Temporal aspects}

\begin{figure}[h]
\centering
\includegraphics[width=0.9\linewidth]{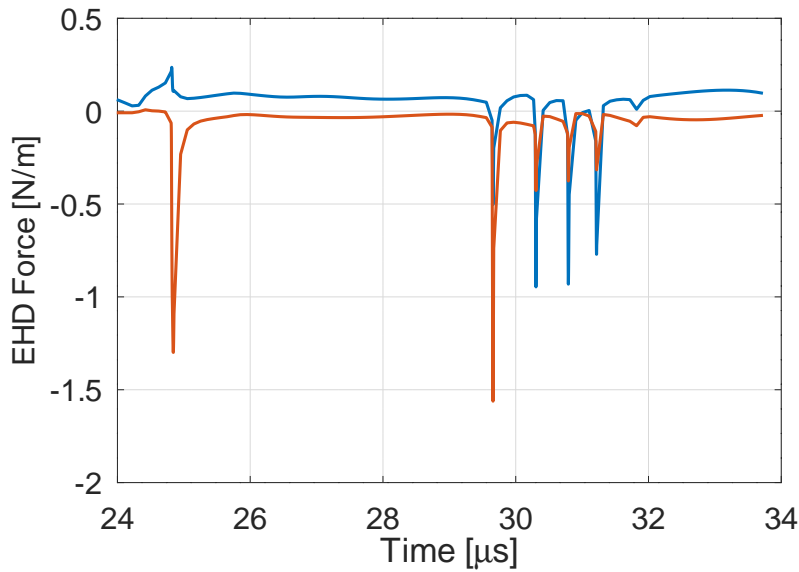}
\caption{Space integrated EHD force (x and y components) [N/m] vs Time [$\mu$s] inside a full AC cycle (3rd period- see Ref.~\cite{jointpaper}) }
 \label{fig:force1}
 \end{figure}

 The space-integrated EHD force (x and y component) versus time (during the third period of actuation) extracted from the detailed simulations of Ref~\cite{jointpaper}, is shown in Fig.~\ref{fig:force1}. The effects of both the streamer and micro-discharges are quite remarkable : The x-force is strongly negative during each micro-discharge formation while it becomes positive in the relaxation phase (thin layer propagation) during the negative going cycle. In the positive going cycle the x-force is always positive and the streamer produces a pulse of positive force. The streamer discharge seems to have a very important influence on the y-force : While the y-force remains a lot weaker that its x-component, each streamer produces a significant negative y-force. 
 
 This suggests that the definition of the EHD forcing as a push-push or push-pull action is misleading. Both phases contribute positively to the EHD x-directed force but strong negative parts exists during the negative phase too. As we will see below at the spatial distribution and resulting ionic wind profiles, the negative parts can form zones of strong negative flow (at least under the operational conditions studied here). Moreover, the y-directed forcing is negative throughout the AC cycle. We consider safe to speculate that depending on various operational characteristics (applied voltage, AC frequency, dielectric material, electrode thickness), the AC-DBD might favor the positive or negative parts during the negative phase resulting to push-push or push-pull effects. Thus, the temporal forcing or velocity profiles extracted from experiments should take into account these operational parameters and relevant time-scales as well as the measurement locations for both velocity components. 
 
\subsection{EHD force distribution - Spatial aspects}
As the fluid response takes place in much longer time-scales than these two phases and the AC frequency of operation, the time-averaged EHD-force provides a good representation of the continuous EHD forcing and resulting flow. The time-averaged force has been calculated during an AC period (third period of Ref.~\cite{jointpaper}) and its spatial distribution (magnitude, x and y components) is plotted in Fig.~\ref{fig:force2}. The EHD force occupies a volume of approx. 4-5 mm in x-direction and 1.5-2 mm in y-direction. It is very high in a small volume near the exposed electrode where its mainly negative-directed (both in x and y directions). The X-directed EHD force is positive and important inside three regions. The first is linked to the initiation of the streamer discharge - the zone at a distance of approx. 0.5 mm where strong ionization occurs. The second is a zone very close to the dielectric layer where negative ions exist during the negative phase. The third is the zone in front of the streamer final elongation length linked to its propagation and dielectric charging during the positive phase. It is thus obvious that the positive going cycle (streamer regime) has important implications to the EHD force production and especially to its spatial distribution in both phases. The negative x-directed EHD force zone near the exposed electrode is linked to the cathode layer formation during the negative phase. This zone is also very important and simplified models ofter neglect it. The negative y-directed EHD force is located mainly close to the active electrode and a layer attached to the dielectric. This zone is due to both phases (positive charges accelerated into the streamer sheath during the positive phase and negative ion charging drifting towards the dielectric as the thin layer moves downstream during the negative phase). The reader should refer to Ref.~\cite{jointpaper}) for more details on the discharge evolution in each AC subcycle which support all of the above. The total elongation of the EHD force is approx. 4 mm. To our knowledge this is the first time that such a result has been obtained - one that links the EHD force distribution with the streamer regime and demonstrates the experimentally observed elongation of the EHD forcing (see Ref.~\cite{benard2017highly} for example). We also note that a longer streamer elongation (due to photoionization effects or streamer pulse repetition during the positive phase under lower actuation frequency) should reproduce similar effects and elongate the EHD force localization even more. As streamers have been experimentally observed to propagate at distances in the order of 10 mm, our results should translate to such cases too. 

  \begin{figure}[h]
\centering
\includegraphics[width=1\linewidth]{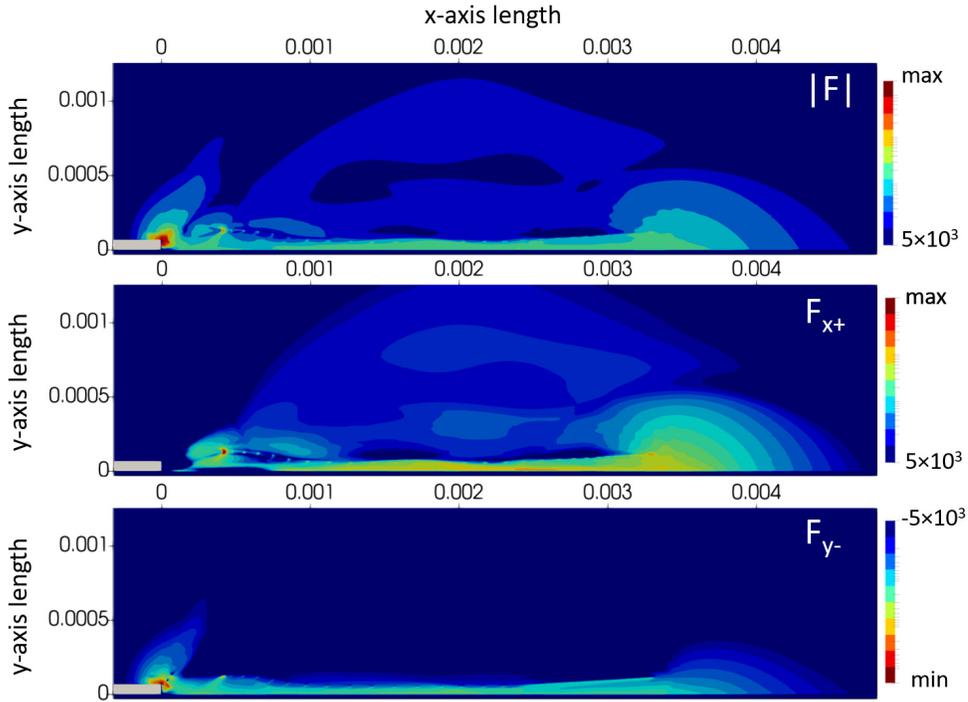}
\caption{Time-integrated (over 1 AC period) EHD force distribution - Force magnitude and $F_{x}$ (positive part), $F_{y}$ (negative part) components  [N/$m^{3}$, log-scale - scaled to min of 1000 N/$m^{3}$ (or max of -1000 N/$m^{3}$ for the negative parts) for visualization purposes]. Note that for the negative $F_{y}$ part we have used the expression $sgn(F_{y})lg(abs(F_{y}))$ in order to represent the spatial distribution in a log scale. The maximum length of the EHD force zone ($\approx$4 mm) coincides with the maximum elongation length of the streamer discharge during the positive phase. Max (min) values are: $|F|= 1.6\times 10^{7}, F_{x}=1.7\times 10^{6}, F_{y}=-1.2\times 10^{7}$ N/$m^{3}$.}
 \label{fig:force2}
 \end{figure}

  \begin{figure}[h]
\centering
\includegraphics[width=1\linewidth]{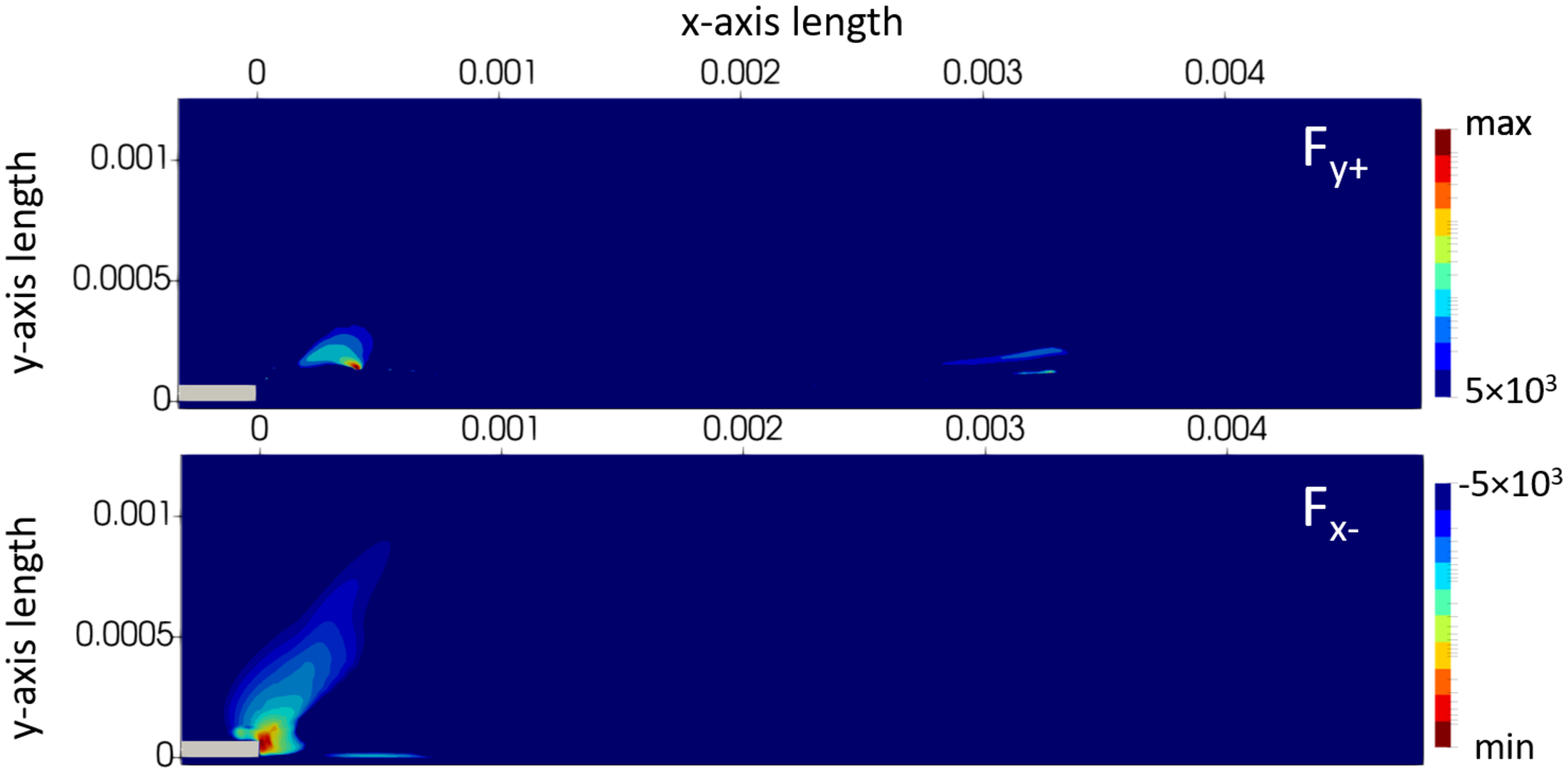}
\caption{Time-integrated (over 1 AC period) EHD force distribution - $F_{x}$ (positive part), $F_{y}$ (negative part) components  [N/$m^{3}$, log-scale - scaled to min of 1000 N/$m^{3}$ (or max of -1000 N/$m^{3}$ for the negative parts) for visualization purposes]. Note that for the negative $F_{x}$ part we have used the expression $sgn(F_{x})lg(abs(F_{x}))$ in order to represent the spatial distribution in a log scale. The strong negative x-force near the exposed electrode is apparent. Max (min) values are: $F_{x}=-1.4\times 10^{7}, F_{y}=5.4\times 10^{5}$ N/$m^{3}$. }
 \label{fig:force2}
 \end{figure}

\subsection{Ionic wind spatial profiles}
\begin{figure}[h]
\centering
\includegraphics[width=1\linewidth]{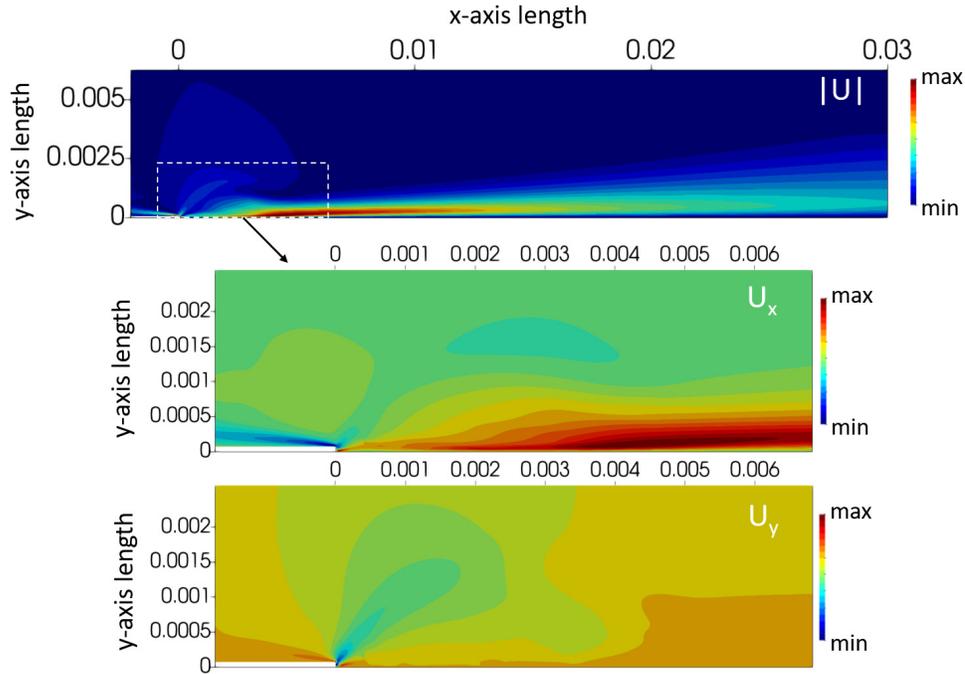}
\caption{Steady state flow field - Velocity magnitude contours [m/s] and zoom near the HV electrode zone. Min value is 0 m/s, Max value is 15.2 m/s. The formation of the wall jet and position of maximum positive velocity coincide with the maximum elongation length of the streamer discharge and the enhanced field downstream during the positive phase.}
 \label{fig:flow1}
 \end{figure}

\begin{figure}[h]
\centering
\includegraphics[width=0.9\linewidth]{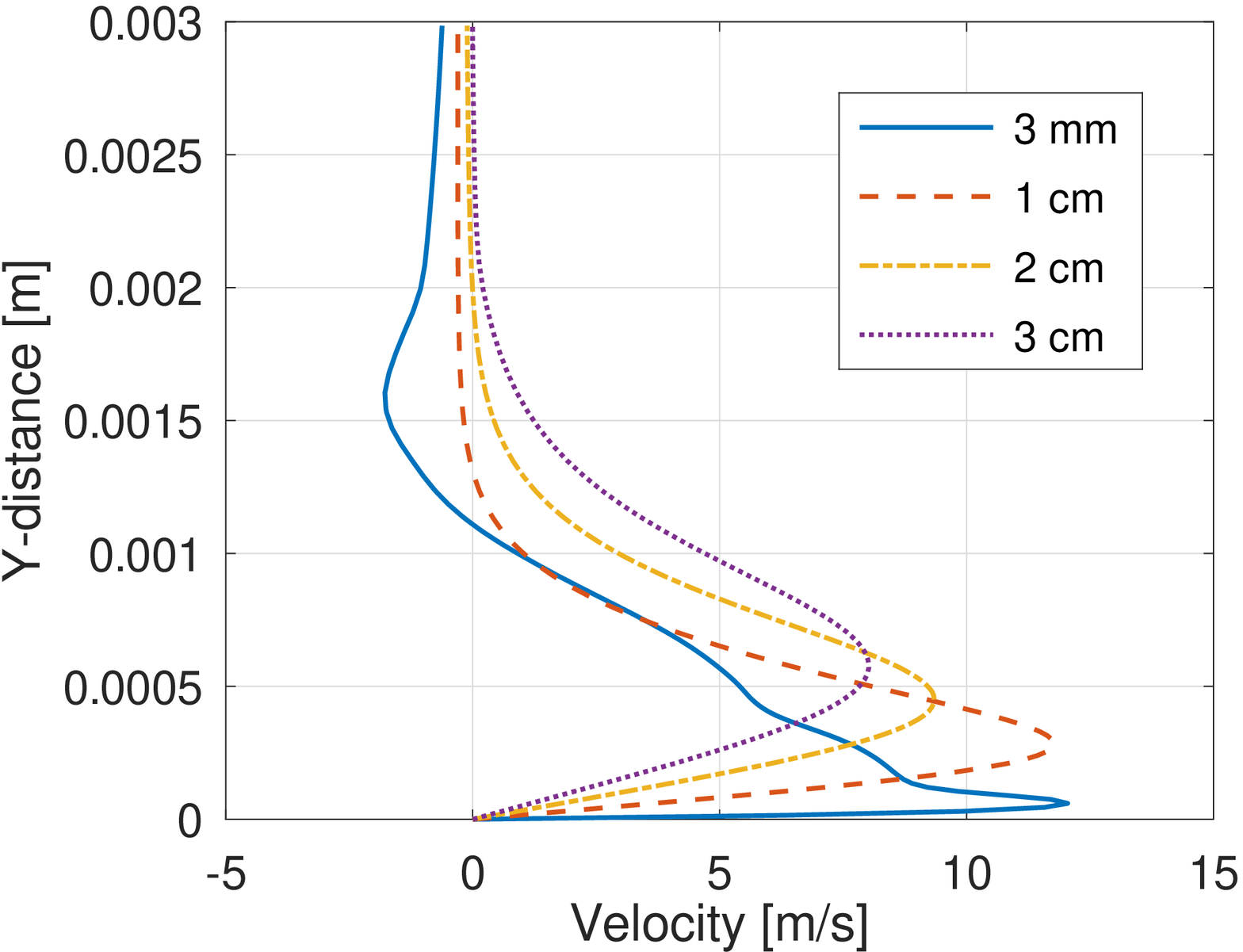}
\caption{Velocity profiles at 3 mm, 1 cm, 2 cm and 3 cm from the HV electrode }
 \label{fig:flow2}
 \end{figure}

The time-averaged EHD body force term has been incorporated into a CFD solver (openFOAM~\cite{weller1998tensorial}) in order to calculate the flow field resulting from the AC-DBD actuators (Fig.~\ref{fig:flow1}). We note here that the total time-averaged space-integrated x-directed force is 64 mN/m while the total y-directed force is -45 mN/m. Menter's k-$\omega$ Shear Stress Transport (SST) turbulence model~\cite{menter1994two} is used for solving the Reynolds-averaged Navier–Stokes equations (RANS) which has already been proven adequate for surface AC-DBD induced wall-jet flows~\cite{kourtzanidis2014modelisation, kourtzanidis2013numerical}. Fig.~\ref{fig:flow1} presents the steady-state velocity contours while Fig.~\ref{fig:flow2} the corresponding velocity profiles at a distance of 3 mm, 1, 2 and 3 cm from the exposed electrode edge. The wall jet flow reaches maximum speeds at a height of approx. 0.5 mm (for 2 and 3 cm) from the dielectric surface in good agreement to experimentally obtained velocity profiles (see Ref.~\cite{moreau2007airflow} and references therein). The thickness of the boundary layer wall jet ranges from 1-2 mm. The maximum velocity occurs at a distance of approx. 4.5 mm from the exposed electrode as can been seen from the zoomed sub-figures in Fig.~\ref{fig:flow1}. To our knowledge, this result is also novel: Not only it clearly demonstrates the importance of the streamer propagation and subsequent dielectric charging (in both phases as described in Ref.~\cite{jointpaper}) to the ionic wind spatial profile but it also explains the experimental profiles in a physical manner linked to the plasma formation and not purely to fluid dynamics. The maximum elongation length of the EHD force and the resulting (positive) ionic wind maximum at a distance of 4-4.5 mm coincide with the enhanced electric field zone just ahead of the maximum elongation length of the streamer discharge, showcasing its influence on the ionic wind spatial distribution. We note here that Ref.~\cite{moreau2018ionic}, Ref.~\cite{defoort2020ionic} and Ref.~\cite{sato2017enhanced} point out towards this direction too. We also note that the maximum elongation distance of the streamer discharge is subject to various parameters (AC frequency, applied voltage, dielectric constant and thickness) and thus the maximum of the ionic wind can be found quite further downstream for different test cases. This result is in very good agreement with the experimental profiles retrieved in various studies~\cite{zhang2020flow, debien2012unsteady, durscher2012evaluation, kotsonis2012performance} which indicated that the maximum velocity occurs at a distance of several millimeters downstream the exposed electrode and could not been replicated so far by numerical experiments. In addition, a negative flow region is observed initiated near the exposed electrode - limited in volume compared to the positive flow. This thin jet flow, induced by the strongly negative EHD zone near the electrode as we have seen, might indicate that opposing flows are present in DBD actuators - another aspect that needs further investigation and might have been ignored so far, an aspect that surely depends on dimensioning (e.g. electrode thickness) and location of measurements/control volume choice. We note finally that the high velocities obtained compared to the experimental results (as well as the thinner jet profiles) are directly linked to the high AC frequency used in the simulations - the goal of this study is the qualitative explanation of the spatial profiles and discrepancies observed which can be extrapolated to lower frequencies without loss of generality. We have already validated this by scaling the EHD force magnitude and manage to reproduce velocity magnitudes and profiles that agree with experimental values. In any case, the reader is referred to Ref.~\cite{jointpaper} for a detailed analysis of each phase along with charge evolution, electric field and surface charging distributions at different time instants as well as discussion on assumptions made.

\section{Conclusion}
We provided an explanation for the ionic wind spatial profiles induced by surface AC-DBD actuators. Based on a detailed numerical study of the surface AC-DBD actuator (presented in Ref.~\cite{jointpaper}), we demonstrated that the elongation of the EHD force and local maxima of the ionic wind are mainly due to the streamer regime of the positive phase but also the presence of a thin negative ion layer during the negative phase attached to the dielectric (which also links to the the streamer regime). Dielectric charging plays a crucial role on the volumetric charge redistribution and consequent EHD force production zones. A strong negative force region also exists near the exposed electrode linked mostly to the negative phase (micro-discharge formation). Therefore, a push-push or pull-pull scenario should depend on the localization of the measurements as well as the operational characteristics of the actuator. We have proposed a detailed explanation behind the EHD production zones and backed up our claims with numerically extracted profiles of the EHD force and the ionic wind.  Our results indicate that streamers can be used to enhance the EHD force and/or create localized distributions. Apart of the obvious implications to aerodynamic flow control, several domains can leverage such findings to improve EHD flows or create novel devices. Improved actuators can be designed based on repetitive streamer production and subsequent charge drift. The influence of streamers to EHD force production is under study in simplified configurations such as point-to-plane discharges. Such optimized devices could also be used for in-atmosphere propulsion systems replacing typical corona based ionic propulsion systems~\cite{xu2018flight, xu2019higher, xu2019dielectric}. Lastly, our results indicate that while ns-DBDs do not create any significant ionic wind (as the streamer production is terminated by the nanosecond pulse and no strong positive electrostatic field exists after its termination to enhance positive ion drift) superposition of DC fields over the nanosecond pulse could lead to enhanced forcing towards optimized and efficient actuator or in-atmosphere propulsion systems. 

\section*{Acknowledgments}

This work has received funding under the E.U. MSCA-RISE - Marie Skłodowska-Curie Research and Innovation Staff Exchange (RISE)  scheme as part of the project `Control of Turbulent Friction Force' (CTFF) - Grant agreement ID: 777717.
\section*{References}

\bibliography{mybib}
\bibliographystyle{abbrv}
\end{document}